\documentclass[prl,twocolumn,showpacs,letterpaper,showpacs,superscriptaddress]{revtex4}
\usepackage{graphicx,amsmath,amssymb,amsfonts,latexsym,color,dcolumn,bm,epsfig}
\begin{document}

\newcommand{\beq}{\begin{equation}}
\newcommand{\eeq}{  \end{equation}}
\newcommand{\bea}{\begin{eqnarray}}
\newcommand{\eea}{  \end{eqnarray}}
\newcommand{\bit}{\begin{itemize}}
\newcommand{\eit}{  \end{itemize}}

\newcommand{\ket}[1]{\left| #1 \right\rangle}
\newcommand{\bra}[1]{\left\langle #1\right |}

\title{Direct measurement of finite-time disentanglement induced by a reservoir}

\author{M. Fran\c ca Santos}
\affiliation{Instituto de F\'{\i}sica, Universidade Federal do Rio
de Janeiro, Caixa Postal 68.528, 21.941-972, Rio de Janeiro, Brazil}

\author{P. Milman}
\affiliation{Laboratoire de Photophysique Mol\'eculaire du CNRS,
Univ. Paris Sud, B\^atiment 210-Campus d'Orsay, 91405 Orsay Cedex,
France}

\author{L. Davidovich}
\affiliation{Instituto de F\'{\i}sica, Universidade Federal do Rio
de Janeiro, Caixa Postal 68.528, 21.941-972, Rio de Janeiro, Brazil}

\author{N. Zagury}
\affiliation{Instituto de F\'{\i}sica, Universidade
Federal do Rio de Janeiro, Caixa Postal 68.528, 21.941-972, Rio de
Janeiro, Brazil}

\date{\today}

\begin{abstract}

We propose a method for directly probing the dynamics of
disentanglement of an initial two-qubit entangled state, under the
action of a reservoir. We show that it is possible to detect
disentanglement, for experimentally realizable examples of decaying
systems, through the measurement of a single observable, which is
invariant throughout the decay. The systems under consideration may
lead to either finite-time or asymptotic disentanglement. A general
prescription for measuring this observable, which yields an
operational meaning to entanglement measures, is proposed, and
exemplified for cavity quantum electrodynamics and trapped ions.

\end{abstract}


\maketitle

\indent Entanglement is the most characteristic trait of quantum
mechanics~\cite{schrodinger1935}. As such, it has led not only to
intense debate since the beginnings of quantum mechanics~\cite{epr},
but also to a variety of possible applications, ranging from
communications~\cite{bennett,gisin} to computation~\cite{nielsen}.
Decoherence and entanglement are closely connected phenomena: not
only decoherence follows from the entanglement of the system of
interest with the rest of the Universe~\cite{zurek}, but also it is
responsible for the fragility of entanglement in systems interacting
with reservoirs. Because of this, understanding the basic mechanisms
of decoherence and disentanglement has both fundamental and
practical implications. However, even for simple systems of two
qubits, the available entanglement measures, like concurrence as
introduced by Wootters\cite{wootters}, involve mathematical
operations that do not seem to have a direct physical
interpretation. It is desirable, therefore, to find examples of
systems for which an entanglement measure could be associated to an
observable that could be easily measured. This would not only make
it easier to follow the dynamics of disentanglement for systems
interacting with reservoirs, but it could be helpful for the
physical interpretation of this process. Here we show that, for an
experimentally realizable example of a decaying system, it is
possible to directly measure concurrence, through the detection of
an observable that is invariant throughout the evolution of the
system.

A state of a bipartite system is disentangled or separable if it can
be written in the form \beq \hat\rho=\sum_i
p_i\hat\rho_i^A\otimes\hat\rho_i^B,\quad (p_i\ge0)\,.
 \eeq

For a pair of qubits, described by the density operator $\rho$,
entanglement may be quantified by the concurrence\cite{wootters}
\begin{equation}
{\cal C}(\hat\rho)={\rm max}\{0,
\sqrt{\lambda_1}-\sqrt{\lambda_2}-\sqrt{\lambda_3}-\sqrt{\lambda_4}\}\,,
\end{equation}
where the $\lambda_i$'s are the eigenvalues, in decreasing order, of
the Hermitian matrix
$\hat\rho(\hat\sigma_y\otimes\hat\sigma_y)\hat\rho^\ast(\hat\sigma_y\otimes\hat\sigma_y)$.
It ranges from ${\cal C}=0$ for a separable state to ${\cal C}=1$
for a maximally entangled state.

The connection between disentanglement of bipartite systems and
decoherence has been recently analyzed, within the framework of
models that may lead to finite-time disentanglement, while the
individual subsystems decay asymptotically in
time\cite{diosi,dodd,eberly1,andre,eberly2}. These examples are
interesting insofar as they allow a clear separation between the
disentangling and the decay dynamics. In~\cite{eberly1}, a realistic
system was considered, consisting of a pair of two-level atoms, each
interacting with its own reservoir. Depending on the initial atomic
state, one may have either finite-time or asymptotic
disentanglement.

Simple methods to detect this transition from entanglement to
disentanglement would be clearly desirable. However, the
non-physical nature of the operations involved in the definition of
concurrence imply that such a detection is usually a challenging
problem, requiring the full tomographic reconstruction of the state.
Entangled states may also be identified through entanglement
witnesses, which are non-positive operators that are positive in the
subspace of separable states\cite{werner}. Entanglement witnesses
allow one to identify some, but not all, entangled states.

For two trapped ions, entanglement between corresponding two-level
internal states has been established either by means of a witness
involving a single matrix element of the two-qubit density
operator~\cite{wineland2000,blatt2005}, or by measuring the full
concurrence via tomographic reconstruction of the
state~\cite{blatt2004}.

In this paper, we show that it is actually possible to detect,
through the measurement of a single observable, finite-time
disentanglement of an initial pure state of two qubits, which
evolves under the action of a reservoir. This observable is a
``perfect witness'' for the class of states here considered: any
entangled state in this class would lead to a negative value for
this operator, which is precisely equal to minus the concurrence of
the state. Remarkably, the same observable yields the concurrence
throughout the evolution of the system. It is possible therefore, by
measuring this operator as the system evolves, to pinpoint the
precise moment when disentanglement occurs. Also, a simple physical
interpretation can be given to concurrence in this case. Our
proposal is within present experimental capabilities in systems like
trapped ions\cite{wineland2000,blatt2005}, cavity quantum
electrodynamics (cavity QED)\cite{haroche1,haroche2}, circuit
quantum electrodynamics\cite{schon}, and nuclear magnetic
resonance\cite{nmr}.

We consider here initial states of the form
$|\Psi(0)\rangle=|\alpha||00\rangle+|\beta|\exp(i\theta)|11\rangle$,
where $0$ and $1$ correspond to the ground and excited state of each
qubit, respectively. The two qubits may stand for spin-down and up
internal states of two trapped ions, or to one- and zero-photon
states of two modes of the electromagnetic field in the same or in
different cavities. States like this have been realized in
experiments with trapped
ions~\cite{blatt2004,blatt2005,wineland2003,wineland2005}. The phase
$\theta$ is attributed by the preparation process. We assume that
the two qubits evolve under the influence of low-temperature
independent and identical reservoirs, so that higher-energy states
are not populated.

Under these conditions, we show that the concurrence of the density
operator evolving from the above initial state is equal to ${\cal
C}(t)={\rm max}\left\{0,-W(t)\right\}$, where $W(t)={\rm
Tr}\left[\hat\rho(t) \hat W_\theta\right]$, and $\hat
W_\theta=1-2|\Phi(\theta)\rangle\langle\Phi(\theta)|$, with
\begin{equation}\label{bell}
|\Phi(\theta)\rangle=(|00\rangle+e^{i\theta}|11\rangle)/\sqrt{2}\,.
\end{equation}

The observable $\hat W_\theta$ is a ``perfect witness," the same
throughout the evolution of the system: it is positive for separable
states, and negative for entangled states.

We note that $W(t)=1-2P(t)$, where $P(t)$ is the probability of
finding the system in the state $|\Phi(\theta)\rangle$ at time $t$.
It is easy to show that other choices of the relative phase in
(\ref{bell}), different from $\theta$, would yield smaller values of
the corresponding probability. This yields a simple physical
interpretation of concurrence in this case: it is twice the maximal
excess probability, with respect to 50\%, of finding the state of
the system in a maximally entangled state corresponding to the
subspace spanned by $\{|00\rangle,|11\rangle\}$.  One should note
that evaluation of $W(t)$ amounts to measuring this probability. We
show now that this can be done in a simple way, by inverting the
process which yields $|\Phi(\theta)\rangle$ from the state
$|00\rangle$.

State (\ref{bell}) may be obtained from the state $|00\rangle$ by
applying to it an unitary transformation $\hat U(\theta)$ composed
of a one-qubit $\pi/2$ rotation followed by a CNOT operation.
Therefore, the probability $P(t)$ can be written as
\begin{equation}\label{inverse}
P(t)=\langle\Phi(\theta)|\hat\rho(t)|\Phi(\theta)\rangle=\langle00|\hat
U^{-1}(\theta)\hat\rho \hat U(\theta)|00\rangle\,,
\end{equation}
that is, the probability of finding the system in the state
(\ref{bell}) can be obtained by applying to the system the inverse
of $\hat U(\theta)$, and then detecting the probability of finding
both qubits in the ground state.

The general format, in the basis
$|11\rangle,|10\rangle,|01\rangle,|00\rangle$, of the time-dependent
density matrix that evolves from state $|\Psi(0)\rangle$ is:
\begin{equation}
\hat\rho(t)=\begin{pmatrix}
                         w(t)&0&0&z(t)\\
                          0&x(t)&0&0\\
                          0&0&x(t)&0\\
                         z^\ast(t)&0&0&y(t)\,
\end{pmatrix}
\label{rhot}
\end{equation}
with $w(t)$, $x(t)$, and $y(t)$ real. Indeed, the incoherent decay
of the state $|11\rangle$ under the action of the reservoir does not
lead to coherence between the states $|10\rangle$ and $|01\rangle$.
Also, the symmetry of the initial state and the equality of the
damping rates for both reservoirs implies that the population of
theses two states is always the same.

The concurrence corresponding to this state is easily determined as
$C(t)={\rm max} \left\{0,2|z(t)|-2x(t)\right\}$. Therefore, the
state is entangled if and only if $|z(t)|>x(t)$.

The probability of finding the system described by (\ref{rhot}) in
the state (\ref{bell}) is given by
\begin{equation}\label{pfi}
P(t)=\left[1-2x(t)+2|z(t)|\right]/2\,,
\end{equation}
and therefore ${\cal C}(t)={\rm max}\left\{0, 2P(t)-1\right\}$, as
anticipated.

For the initial state $|\Psi(0)\rangle$, ${\cal
C}(0)=2|\alpha\beta^*|=2|z(0)|$, which equals one for the maximally
entangled state $(|00\rangle+e^{i\theta}|11\rangle)/\sqrt{2}$. As
time evolves, the state becomes disentangled when $P(t)$ reaches the
value $1/2$.

This result is rather insensitive to small deviations from the above
state. Thus, if the populations of states $|10\rangle$ and
$|01\rangle$ are slightly different (due, for instance, to unequal
decay rates), then if $\epsilon(t)=|x_1(t)-x_2(t)|\ll |z(t)|$ it is
easy to show that the concurrence is still given by the above
expression, up to terms of order $\epsilon^2$. Also, our method can
be extended to any state in the subspace
$\{|00\rangle,|11\rangle\}$.

\begin{figure}[t]
\includegraphics[height=6cm,width=8cm]{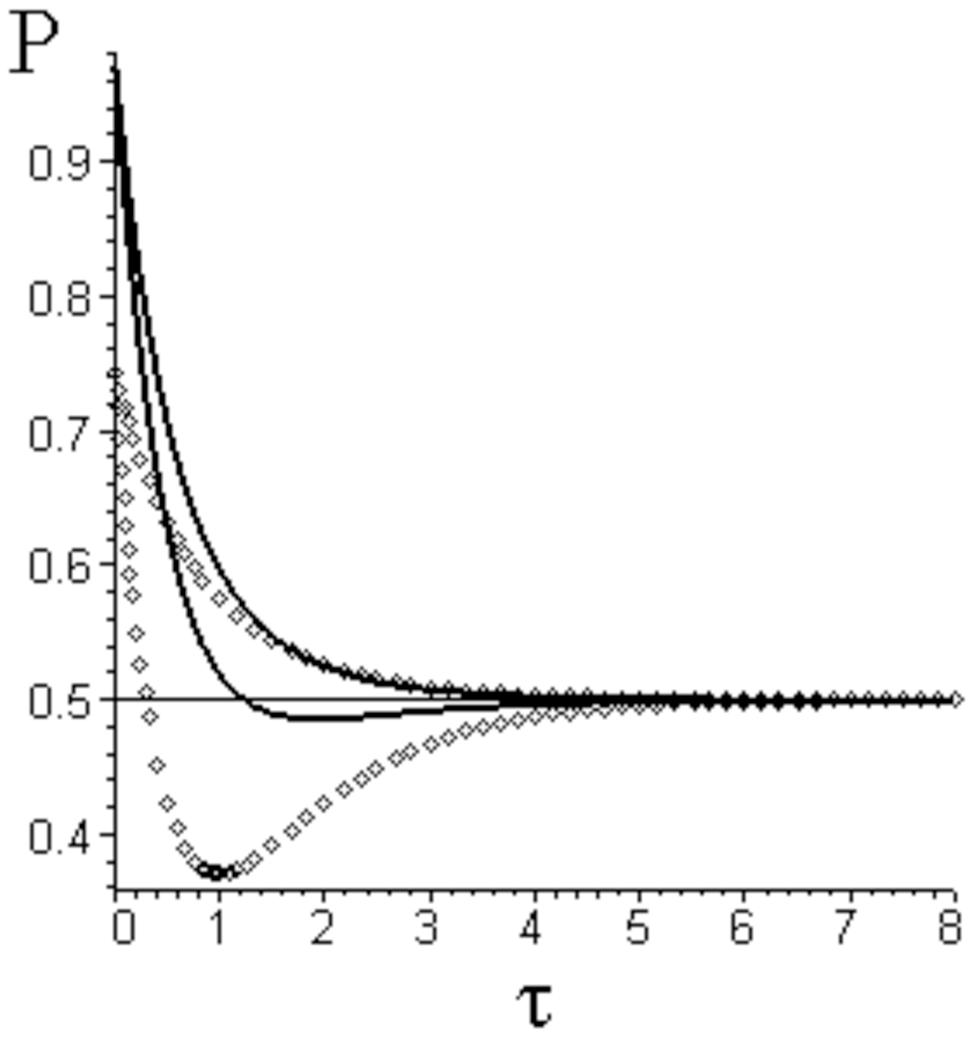}
\caption{$P$ as a function of $\tau = \gamma t$ for four different
initial states, of the form $\alpha|00\rangle+\beta|11\rangle$, with
concurrence ${\cal C}(0)=2|\alpha||\beta|$. Solid thick lines
correspond to ${\cal C}(0)=\sqrt{2}/3$, while dotted lines
correspond to ${\cal C}(0)=\sqrt{15}/16$. Concurrence, as a function
of time, is given by ${\rm max}\left\{0, 2P(t)-1\right\}$.
Entanglement persists as long as $P>1/2$ (this bound corresponds to
the thin solid line). States with the same initial entanglement
exhibit finite-time disentanglement for $|\beta|>|\alpha|$ and
infinite-time disentanglement for $|\alpha|>|\beta|$} \label{fig1}
\end{figure}

We show now that, for the usual linear coupling with markoffian
reservoirs, ${\cal C}(t)$ may reach zero at finite times. The master
equation governing the time behavior of the system can be written in
the Lindblad form~\cite{lindblad}:
\begin{equation}\label{master}
\dot{\hat \rho }= \sum_i (\gamma_i/2)\left ( 2 \hat c_i \hat \rho \hat
c^{\dagger}_i-\hat c^{\dagger}_i \hat c_i \hat \rho-\hat \rho \hat
c^{\dagger}_i \hat c_i \right ),
\end{equation}
where $\gamma_i$ are decay rates and $\hat c_i, \hat c^{\dagger}_i$
are operators representing the coupling to the reservoir. For atoms
decaying spontaneously due to the coupling to a zero temperature
reservoir, $\hat c=\hat \sigma_-$, where $\hat\sigma_-$ is the
lowering operator acting on the electronic levels of the atom. For
modes of the electromagnetic field in one or two cavities, coupled
to a zero- temperature reservoir, the decaying dynamics is analogous
to the ionic system, provided that the photonic state is initially
in the subspace $\{\ket{0},\ket{1}\}$. In this case, $\hat c_i$
correspond to annihilation operators for photons, for each of the
two cavity modes. The solution of (\ref{master}) for any initial
state lying in the subspace $\{\ket{00},\ket{11}\}$ is of the form
(\ref{rhot}) with:
\begin{eqnarray}
&&w(t)=w(0)e^{-2\gamma t}, \nonumber \\
&&x(t)=w(0)(1-e^{-\gamma t})e^{-\gamma t}, \\
&&y(t)=y(0)+w(0)(1-e^{-\gamma t})^2,\nonumber \\
&&z(t)=z(0)e^{- \gamma t}. \nonumber
\end{eqnarray}
Consequently, an initial state of the form $\alpha \ket{00}+ \beta
\ket{11}$  becomes separable for $t_s=-\frac{1}{\gamma}\rm
ln(1-\frac{|\alpha|}{|\beta|})$. It is clear that only for states
with $|\beta|>|\alpha|$ is this condition fulfilled for finite
times. The probability $P(t)$  is shown in Fig.~\ref{fig1}, for four
different initial states. One should note that states with the same
initial entanglement exhibit finite-time disentanglement for
$|\beta|>|\alpha|$ and infinite-time disentanglement for
$|\alpha|>|\beta|$.
For states of the form $\gamma|01\rangle+\delta|10\rangle$,
disentanglement occurs asymptotically in time. This justifies our
interest in the special class of states here considered, since they
allow a clear separation between the processes of disentanglement
and decoherence, besides allowing a simple monitoring procedure of
the disentanglement process.

One should note that the method of measurement outlined above is
valid only if the operations on the qubits do not involve auxiliary
systems, as is the case for instance in nuclear magnetic resonance.
On the other hand, the realization of quantum gates in ion traps
(where the vibrational mode is used for the realization of a CNOT
gate), or in cavity QED (where the cavity mode mediates the
interaction between two atoms, or an atom mediates the interaction
between two cavity modes) require auxiliary systems. In these cases,
the operation $\hat U(\theta)$ does not depend only on the qubit
operators, so that Eq.~(\ref{inverse}) does not hold. In spite of
this, the above method can still be used to measure entanglement,
the results differing from the concurrence by a scaling constant. We
exemplify this with application to ion traps and cavity QED.

For two trapped ions, the production of the initial state and the
measurement of $P(t)$ can be done by employing standard techniques
involved in the production and detection of entangled
states~\cite{wineland2003,wineland2005,blatt2005}.

In these experiments, maximally entangled states are produced through a combination of red-sideband and
blue-sideband transitions and detected using the shelving technique: the ion  undergoes cycling transitions
to an excited unstable state if and only if it is in the ground state, the resulting fluorescence being
measured with very high detection efficiency (of the order of $99\%$). The fluorescence yield of each ion
is thus proportional to the probability of finding it in the ground state.

One should note that the decay of the ions does not lead to
population of the vibrational modes, which remain in the ground
state. We consider the initial state
$(|\alpha||g_1g_20\rangle+e^{i\theta}|\beta||e_1e_20\rangle)/\sqrt{2}$,
where $e_i$ and $g_i$ stand respectively for the upper and lower
internal states of the ions ($i=1,2$) and integer numbers stand for
the states of the vibrational mode. Disentanglement of this state
can be measured by applying to the first ion a red-sideband pulse,
so that $\ket{g_11}\rightarrow\ket{e_10}$ and
$\ket{e_10}\rightarrow-\ket{g_11}$.  This operation transforms the
state $\hat\rho(t)\otimes\ket{0}\bra{0}$, with $\hat\rho(t)$ given
by (\ref{rhot}), into
\begin{eqnarray*}
&&w(t)\ket{g_1e_21}\bra{g_1e_21}+y(t)\ket{g_1g_20}\bra{g_1g_20}\\
&&-z(t)\ket{g_1e_21}\bra{g_1g_20}-z^*(t)\ket{g_1g_20}\bra{g_1e_21}\\
&&+x(t)(\ket{g_1e_20}\bra{g_1e_20}+\ket{g_1g_21}\bra{g_1g_21})\,.
\end{eqnarray*}

Next a blue-sideband pulse is applied on the second ion, so that
$\ket{g_20}\rightarrow(\ket{g_20}-\exp(i\delta)\ket{e_21})/\sqrt{2}$,
$\ket{e_21}\rightarrow(\exp(i\delta)\ket{g_20}+\ket{e_21})/\sqrt{2}$,
and
$\ket{g_21}\rightarrow\cos(\pi\sqrt{2}/4)\ket{g_21}-\sin(\pi\sqrt{2}/4)\exp(-i\delta)\ket{e_22}$.
The probability $P_{gg}(\delta,t)$ that both ions are in the ground
state is
\begin{equation}\label{qed}
P_{gg}(\delta,t)={1}/{2}-|z(t)|\cos(\theta-\delta)-\eta x(t)\,,
\end{equation}
where $\eta=\sin^2(\pi\sqrt{2}/4)$. Choosing $\delta$ so that
$\cos(\theta-\delta)=-\eta$ one gets, calling ${\cal P}_{gg}(t)$ the
value of $P_{gg}(\delta,t)$ at this point:
\begin{equation}
2{\cal P}_{gg}(t)-1=\eta\left[2P(t)-1\right]\,,
\end{equation}
with $P(t)$ given by (\ref{pfi}). This shows that ${\rm
max}\{0,2{\cal P}_{gg}-1\}$ is proportional to the concurrence. One
should note that even if the initial phase $\theta$ is not known,
measurement of $P_{gg}(\delta,t)$ for three values of the phase
$\delta$, leading to three linearly independent equations for
$x(t)$, $|z(t)|$, and $\theta$, would allow one to determine the
concurrence.

\begin{figure}[t]
\includegraphics[height=3cm,width=8cm]{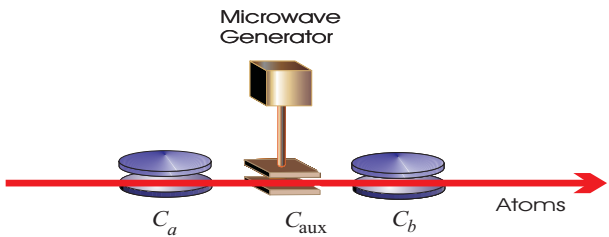}
\caption{Experimental setup for probing finite-time decoherence in
cavity QED.} \label{fig2}
\end{figure}

For cavity QED, the experimental setup involves two high-$Q$
cavities ($C_a$ and $C_b$) on either side of a low-$Q$ cavity
($C_{\rm aux}$), as shown in Fig.~\ref{fig2}. A two-level atom
crosses this system, interacting resonantly with the three cavities.
The field in $C_{\rm aux}$ can be taken as classical, generating a
$\pi$ rotation of the atomic state. The fields in $C_a$ and $C_b$
are initially in the vacuum state. The interaction time between the
atom and each cavity is adjusted, for open-cavity
geometry\cite{haroche1,haroche2}, by Stark-shifting the relevant
atomic levels, so as to tune them into resonance with the cavity
mode for the proper amount of time.

The first step is the preparation of the entangled two-cavity state.
This is done by sending a two-level atom, initially in the excited
state $|e\rangle$, through the three cavities. The interaction time
with the first cavity ($C_a$) is adjusted so that state
$|\Psi_1\rangle=\alpha|e,0,0\rangle-\beta|g,1,0\rangle$ is produced
($|i,n,m\rangle$ denotes atomic state $|i\rangle$ and Fock states $
|n,m\rangle$ for cavities $C_a$ and $C_b$ respectively). That is,
the atom has a probability $|\alpha|^2$ of remaining in the initial
state, and a probability $|\beta|^2$ of decaying to the state
$|g\rangle$, leaving one photon in the resonant mode of $C_a$. The
atomic $\pi$ rotation in $C_{\rm aux}$, takes $|\Psi_1\rangle$ into
$|\Psi_2\rangle=\alpha|g,0,0\rangle+\beta|e,1,0\rangle$, and another
$\pi$ rotation, this time with the resonant mode of $C_b$, produces
the desired entangled state
$|\Psi_f\rangle=\alpha|0,0\rangle+\beta|1,1\rangle$ of the two
modes. The atom leaves the setup always in the ground state.

Due to cavity losses, the prepared state $|\Psi_f\rangle$ evolves
into the mixed state described by Eq.~(\ref{rhot}). In order to
measure the disentanglement of the state of the high-$Q$ cavities,
another two-level atom is sent again through the setup. This atom
enters $C_a$ in the ground state and undergoes a Rabi $\pi$ rotation
if there is one photon in the resonant mode. Next, $C_{\rm aux}$ is
used to generate yet another atomic $\pi$ rotation and then the atom
state undergoes a final rotation in $C_b$, so that
$\ket{g1}\rightarrow(\ket{g1}+\ket{e0})/\sqrt{2}$,
$\ket{e0}\rightarrow(-\ket{g1}+\ket{e0})/\sqrt{2}$. The rotation in
$C_{\rm aux}$ includes a control phase $\delta/2$ ($|g\rangle
\rightarrow e^{-i\delta/2}|e\rangle$ and $|e\rangle \rightarrow
-e^{i\delta/2} |g\rangle$). Finally, the atomic internal state is
measured by ionization. The probability $P_e(\delta,t)$ of finding
the atom in the excited state is given by the same expression
(\ref{qed}) obtained for the two-ion system, leading to the
concurrence in the same way.

Finite-time separability is related, in the example discussed above,
to properties of the reservoir and the initial state. The same
phenomenon occurs for a diffusive reservoir, acting on the two-qubit
system. The corresponding contribution for the master equation can
be written in the form~(\ref{master}) with $\hat c_1=\hat \sigma_-$
and $\hat c_2=\hat \sigma_+$ for each reservoir. Due to the symmetry
of this reservoir, all states belonging to the subspace
$\{\ket{00},\ket{11}\}$, or yet $\{\ket{01},\ket{10}\}$, display
finite-time separability, independently of the relative population
of the states. This is consistent with the findings in
\cite{eberly2}, which considered the action of classical noise on a
two-qubit system.

Since the work of John Bell~\cite{bell} the subtle property of
entanglement has been subjected to many quantitative tests and has
led to intriguing consequences. Understanding the physical meaning
of entanglement measures remains however a major challenge. In this
paper, we have analyzed an example of a decaying two-qubit system
for which it is possible to attribute an operational meaning to an
entanglement measure, valid throughout the decay process, and we
have proposed an experimental procedure which amounts to a direct
measurement of entanglement.

We acknowledge support of the Millennium Institute for Quantum
Information, FAPERJ and CNPq.


\end{document}